\documentstyle[12pt]{article}
\def\appendix#1{
\addtocounter{section}{1}
\setcounter{equation}{0}
\renewcommand{\thesection}{\Alph{section}}
\section*{Appendix \thesection\protect\indent #1}
\addcontentsline{toc}{section}{Appendix \thesection\ \ \ #1}
}
\newcommand{\tr}[1]{\,{\rm tr}\,#1\,}

\def\e{\varepsilon}

\def\be{\begin{equation}}
\def\la{\label}
\def\ee{\end{equation}}
\def\bea{\begin{eqnarray}}
\def\eea{\end{eqnarray}}
\def\eps{\varepsilon}
\def\a{\alpha}
\def\b{\beta}

\def\n{\nabla}
\def\S{\Sigma}
\def\D{\Delta}
\def\G{\Gamma}
\def\g{\gamma}
\def\d{\delta}
\def\o{\omega}

\def\l{\left(}
\def\r{\right)}
\def\p{\partial}
\def\x{\vec{x}}
\def\y{\vec{y}}
\def\z{\vec{z}}

\begin{document}
\title{\hfill{LMU-TPW 99-12} \\
\hfill{UAHEP993} \\
\hfill{hep-th/9907085} \\
\vspace{1cm}
Some Cubic Couplings in Type IIB Supergravity on $AdS_5\times S^5$
and Three-point Functions in SYM$_4$ at Large $N$}
\author{G.Arutyunov$^{a\, c}$
\thanks{arut@theorie.physik.uni-muenchen.de} \mbox{}
 and \mbox{} S.Frolov$^{b\,c}$\thanks{frolov@bama.ua.edu 
\newline
$~~~~~$$^c$On leave of absence from 
Steklov Mathematical Institute,Gubkin str.8, GSP-1, 117966, Moscow, Russia
}
\vspace{0.4cm} \mbox{} \\
$^a$ Sektion Physik,
\vspace{-0.1cm} \mbox{} \\
Munich University
\vspace{-0.1cm} \mbox{} \\
Theresienstr. 37,
\vspace{-0.1cm} \mbox{} \\
D-80333 Munich, Germany
\vspace{0.4cm} \mbox{} \\
$^b$Department of Physics and Astronomy,
\vspace{-0.1cm} \mbox{} \\
University of Alabama, Box 870324,
\vspace{-0.1cm} \mbox{} \\
Tuscaloosa, Alabama 35487-0324, USA
\mbox{}
}
\date {}
\maketitle
\begin{abstract}
All cubic couplings in type IIB supergravity 
on $AdS_5\times S^5$ that involve two scalar fields 
$s^I$ that are mixtures of the five form 
field strength on $S^5$ and the trace of 
the graviton on $S^5$ are derived by using the covariant 
equations of motion and the quadratic action for 
type IIB supergravity on $AdS_5\times S^5$. 
All corresponding three-point functions in SYM$_4$ 
are calculated in the supergravity approximation.
It is pointed out that the scalars $s^I$ correspond not 
to the chiral primary operators in the ${\cal N}=4$ SYM 
but rather to a proper extension of the operators.
\end{abstract}
\newpage
\section{Introduction}
According to the AdS/CFT correspondence \cite{M,GKP,W}, the generating
functional of Green functions in $D=4$, ${\cal N}=4$
supersymmetric Yang-Mills theory (SYM$_4$)
at large $N$ and at strong 't Hooft coupling $\lambda$ coincides with 
the on-shell value of the 
type IIB supergravity action on  $AdS_5\times S^5$. For this reason,
to calculate an $n$-point Green function one has to know
the supergravity action up to the 
$n$-th order. In particular, the normalization constants 
of two- and three-point Green functions \cite{AV}-\cite{LT2} 
are determined 
by the quadratic and cubic actions for physical 
fields of supergravity. 

The particle spectrum of type IIB supergravity on $AdS_5\times S^5$
\cite{KRN,GM} contains scalar fields 
$s^I$ that are mixtures of the five form field strength on $S^5$ and the 
trace of the graviton on $S^5$. The transformation properties of the 
scalars 
with respect to the superconformal group of SYM$_4$ allow 
one to conclude 
that they correspond to chiral primary operators (CPOs) of SYM$_4$.
In \cite{LMRS} the quadratic and cubic actions for the scalars $s^I$ 
have been found and used to calculate all three-point functions of 
normalized CPOs. These three-point functions 
appeared to coincide with the three-point functions of CPOs computed in 
free field theory for generic 
values of conformal dimensions of CPOs.
However, there is an apparent 
contradiction. As was noted in \cite{HF1} (see also \cite{LT2}) 
a three-point function of CPOs calculated in the AdS/CFT 
framework vanishes, if the sum of conformal 
dimensions of any of the two operators equals the 
conformal dimension of 
the third operator, because of the vanishing of the 
cubic couplings of 
the corresponding scalar fields.
Thus we are forced to conclude that the scalars $s^I$ 
used in \cite{LMRS} 
cannot correspond to CPOs.
Another way to come to the conclusion is that the scalars from 
\cite{LMRS}
do not coincide with the original scalars that are mixtures of the 
five-form 
and the graviton but depend nonlinearly on the original 
scalars and their
derivatives. Thus the scalars used in \cite{LMRS} 
do not transform with 
respect to the superconformal group in a proper way 
and cannot correspond to CPOs. 

In this paper we show that a scalar $s^I$ used in 
\cite{LMRS} corresponds to an operator 
which is the sum of a CPO and non-chiral composite operators. The 
non-chiral 
operators are normal-ordered products of CPOs and their descendants,
i.e. so-called double- and multi-trace operators. 

The knowledge of correlation functions of the 
chiral primary operators 
allows one to compute correlation functions of 
all their descendants, in particular, 
the correlation functions of the stress energy 
tensor and $R$-symmetry currents. 
To compute four-point functions\footnote{Some results on four-point 
functions have been obtained in 
\cite{HF1}-\cite{BKRS}.} 
of the chiral operators one has to know the $s^I$-dependent quartic 
terms and all 
cubic terms that involve two scalar fields $s^I$. 
In the present paper, 
as the first step in this direction, we determine 
all such cubic terms. 
It is sufficient to consider only the sector of 
type IIB supergravity 
that depends on the graviton and the four-form potential.
There are four different types of vertices describing interaction 
of two scalars $s^I$
with symmetric tensor fields of the second rank coming 
from the $AdS_5$ components
of the graviton, with vector fields, with scalar 
fields coming from the $S^5$ 
components of the graviton, and with 
scalar fields $t^I$ that are mixtures of the 
trace of the graviton on the sphere and the five 
form field strength on the sphere.

To this end we apply an approach similar to the 
one used in \cite{LMRS}. Namely, we use the quadratic 
action for type IIB supergravity on 
$AdS_5\times S^5$ recently obtained in \cite{AF3} and the 
covariant equations of motion of \cite{S,SW,HW}. Just as 
it was in the case of cubic
couplings of three scalars $s^I$ \cite{LMRS}, to get rid of 
higher-derivative
terms we will have to redefine the original 
gravity fields. Thus the fields entering the final action 
correspond not to descendants of CPOs but to 
extended operators involving  products of CPOs 
and their descendants. However, we expect that for generic 
values of conformal dimensions of these operators, 
their three-point functions coincide with the three-point 
functions of the corresponding descendants
of CPOs. Let us note in passing that the only way to 
find an action 
depending on the fields that correspond
directly to CPOs and their descendants seems to be to derive the  
action starting 
from the covariant action
of \cite{ALS,ALT}. In this way one probably should obtain a 
nonvanishing cubic couplings of scalars $s^I$ corresponding 
to CPOs whose conformal dimensions satisfy the 
relation $\Delta_1 +\Delta_2=\Delta_3$. 
These cubic terms
seem to be of the form suggested in \cite{HF1}.
Unfortunately, the lack of covariance of 
the gauge-fixed
action of \cite{ALS,ALT} makes the analysis extremely complicated. 

The paper is organized as follows. In section 2 we suggest the 
operators that correspond to the scalars $s^I$ from \cite{LMRS}. 
In section 3 we recall
equations of motion for the graviton and the four-form potential, 
and the quadratic actions for the fields under consideration,  
and introduce
notations. In section 4 we obtain cubic couplings of two scalars 
$s^I$ with a
scalar $t^I$, and  with scalars $\phi^I$
coming from the graviton on the sphere, and calculate their 
three-point functions by using results obtained in \cite{FMMR}.
In section 5 cubic couplings of two 
scalars $s^I$ with symmetric second rank tensor 
fields are derived and the corresponding three-point 
functions are found. 
In section 6 we obtain cubic vertices of two scalars
$s^I$ and a vector field, and calculate their 
three-point functions.
Note that three-point functions
of two scalars with a massive vector field, or a massive 
symmetric second rank tensor, were not considered in the 
literature before. In the Conclusion we discuss the results
obtained, and open problems. In the Appendix we recall the 
definitions of scalar, vector and tensor spherical harmonics.
 
\section{Extended chiral primary operators}
In this section we recall the definition of chiral primary operators and introduce a notion of extended chiral primary operators.

According to \cite{LMRS}, CPOs have the form
\bea
O^I(\x )=\frac{(2\pi )^k}{\sqrt{k\lambda^k}}C^I_{i_1\cdots i_k}
\tr (\phi^{i_1}(\x )\cdots \phi^{i_k}(\x )),
\la{cpo}
\eea
where $C^I_{i_1\cdots i_k}$ are totally symmetric traceless rank $k$ 
orthonormal tensors of $SO(6)$: 
$\langle C^IC^J\rangle =C^I_{i_1\cdots i_k}C^J_{i_1\cdots i_k}=\delta^{IJ}$, 
and $\phi^{i}$ are scalars of SYM$_4$. 

The two- and three-point functions of CPOs computed in free theory 
are \cite{LMRS}
\bea
&&\langle O^I(\x )O^J(\y )\rangle =\frac{\delta^{IJ}}{|\x -\y |^{2k}},
\la{cpo2}\\
&&\langle O^{I_1}(\x )O^{I_2}(\y )O^{I_3}(\z )\rangle =\frac 1N
\frac{\sqrt{k_1k_2k_3}\langle C^{I_1}C^{I_2}C^{I_3}\rangle }
{|\x -\y |^{2\a_3}|\y -\z |^{2\a_1}|\z -\x |^{2\a_2}},
\la{cpo3}
\eea
where $\a_i =\frac 12 (k_j+k_l-k_i)$, $j\neq l\neq i$, and 
$\langle
C^{I_1}C^{I_2}C^{I_3}\rangle $ is the unique $SO(6)$ invariant obtained 
by contracting $\a_1$ indices between $C^{I_2}$ and  $C^{I_3}$, 
$\a_2$ indices between $C^{I_3}$ and  $C^{I_1}$, and
$\a_3$ indices between $C^{I_2}$ and  $C^{I_1}$.

According to the AdS/CFT conjecture, there should exist 
fields of type IIB 
supergravity on $AdS_5\times S^5$ that correspond to CPOs. 
The transformation properties of CPOs and supergravity fields 
with respect to the superconformal
group of SYM$_4$ show that these fields seem to be scalar fields 
$s^I$, that are 
mixtures of the five form field strength on $S^5$ and 
the trace of the 
graviton on $S^5$.\footnote{Strictly speaking this correspondence 
between CPOs and scalars $s^I$ may be valid only at linear order 
in supergravity fields. The reason is that the local supersymmetry
transformations of supergravity fields are nonlinear, and, one should 
expect that the induced superconformal transformations are nonlinear too.
Thus the original gravity fields seem to depend nonlinearly on fields
with the linear transformation law.}
To calculate the three-point functions of CPOs in the framework of 
the AdS/CFT correspondence the quadratic and cubic actions for 
the scalars $s^I$ were found in \cite{LMRS}. Then, it was shown that 
for generic values of conformal dimensions of CPOs the normalized
three-point functions computed using the actions precisely coincide 
with the free field theory result (\ref{cpo3}). On the other hand,
as was pointed out in \cite{HF1} the cubic couplings of scalars $s^I$
satisfying one of the three relations:
\bea
k_1+k_2=k_3,\quad k_2+k_3=k_1,\quad k_3+k_1=k_2,
\la{kkk}
\eea
vanish, and, therefore, the three-point functions of the operators
corresponding to scalars $s^I$ vanish too. Thus, scalars $s^I$ used 
in \cite{LMRS} do not correspond to CPOs. We can explain this 
by noting that the scalars $s^I$ from \cite{LMRS} differ from 
the original scalars that are mixtures of the graviton and the 
five-form on $S^5$. The original scalars $s^I$ satisfy equations
which depend on higher-derivative terms. To remove the derivative
terms the following field redefinition was made in \cite{LMRS}
\bea
s^{I_1}= s'^{I_1}+\sum_{I_2,I_3}\left( 
J_{I_1I_2I_3}s'^{I_2}s'^{I_3}+L_{I_1I_2I_3}\n^a s'^{I_2}\n_a s'^{I_3}
\right).
\la{red}
\eea
Namely for the scalars $s'^{I}$ the cubic couplings mentioned 
above vanish. Because of the redefinition (\ref{red}) new scalars 
$s'^{I}$ do not transform with respect to the superconformal group
in a proper way, and, therefore, cannot correspond to CPOs.

From the computational point of view these cubic couplings 
have to vanish because if, say, $k_1+k_2=k_3$ then 
the three-point function (\ref{cpo3}) is nonsingular at $x=y$, but
gravity calculations with a nonvanishing on-shell bulk cubic coupling 
always lead to a function singular at $x=y$, $x=z$ and $y=z$.
By the same reason we expect that $n$-point functions of operators 
corresponding to scalars  $s'^{I}$ (with an additional field 
redefinition which is required to remove higher-derivative terms 
from the $(n-1)$-th order equations of motion for  $s^{I}$) 
would vanish if, say, $k_n=k_1+\cdots +k_{n-1}$. Study of the general 
scalar exchange performed in \cite{HF1} seems to confirm 
the conclusion.

Thus, scalars $s^I$ (here and in what follows we omit the primes on 
redefined fields) correspond to properly extended CPOs which have
vanishing three-point functions if (\ref{kkk}) is fulfilled. Indeed one 
can easily find such an extension of CPOs. Namely, we define 
the extended CPOs  that correspond to scalars  $s^{I}$ as
\bea
\tilde {O}^{I_1}=O^{I_1} -\frac {1}{2N}\sum_{I_2+I_3=I_1}
C^{I_1I_2I_3}O^{I_2}O^{I_3}, 
\la{ecpo}
\eea
where $C^{I_1I_2I_3}=\sqrt{k_1k_2k_3}\langle C^{I_1}C^{I_2}C^{I_3}\rangle $.

It is not difficult to verify that in the large $N$ limit these
operators have the normalized two-point functions (\ref{cpo2}),
the three-point functions (\ref{cpo3}) if (\ref{kkk}) is not satisfied,
and vanishing three-point functions if (\ref{kkk}) takes place.  
However, these operators will require a further modification to be 
consistent with all $n$-point functions computed in the framework of 
the AdS/CFT
correspondence. In general, an extended CPO is the sum of a 
CPO and non-chiral composite operators which are normal-ordered 
products of CPOs and their descendants. Nevertheless, we expect that 
in the large $N$ limit an $n$-point function of extended CPOs  
coincides with $n$-point functions of CPOs for generic values 
of conformal dimensions of the operators. As we will discuss in next 
sections a similar modification is required for operators 
corresponding to other supergravity fields.
\section{Equations of motion and quadratic actions}
To obtain cubic couplings of two scalars $s^I$ with 
other type IIB
supergravity fields it is sufficient to consider only 
the graviton
and the four-form potential. To this end we apply 
the method 
of \cite{LMRS}, and
use the covariant equations of motion \cite{S,SW,HW} and 
the quadratic action for type IIB supergravity on 
$AdS_5\times S^5$
\cite{AF3}. The equations of motion of the 4-form
potential and the graviton are
\bea
F_{M_1...M_5}&=&\frac{1}{5!}\eps_{M_1...M_{10}}F^{M_6...M_{10}},
\la{ffe}\\
R_{MN}&=&\frac{1}{3!}F_{MM_1...M_4}F_N^{~~M_1...M_4}.
\la{gre}
\eea
Here $M,N,\ldots ,=0,1,\ldots 9$ and we use the 
following notations
$$F_{M_1\ldots M_5}=5\partial_{[M_1}A_{M_2\ldots M_5]}=
\partial_{M_1}A_{M_2\ldots M_5}+4~{\mbox{ terms}} ,$$
i.e all antisymmetrizations are with "weight"1. 
The dual forms are defined as
\bea
&&\e_{01\ldots 9}=\sqrt{-G};\quad \
e^{01\ldots 9}=-\frac{1}{\sqrt{-G}}
\nonumber\\
&&\e^{M_1\ldots M_{10}}=
G^{M_1N_1}\cdots G^{M_{10}N_{10}}\e_{N_1\ldots N_{10}}
\nonumber\\
&& (F^*)_{M_1\ldots M_k}=
\frac{1}{k!}\e_{M_1\ldots M_{10}}F^{M_{k+1}\ldots M_{10}}
=\frac{1}{k!}\e^{N_1\ldots N_{10}}G_{M_1N_1}\cdots
G_{M_{k}N_{k}}F_{N_{k+1}\ldots N_{10}}.
\nonumber
\eea
In the units in which the radius of $S^5$ is set to be unity, 
the $AdS_5\times S^5$ background solution looks as
\bea
&&ds^2=\frac{1}{x_0^2}(dx_0^2+\eta_{ij}dx^idx^j)+d\Omega_5^2=
g_{MN}dx^Mdx^N
\nonumber\\
&&R_{abcd}=-g_{ac}g_{bd}+g_{ad}g_{bc};
\quad R_{ab}=-4g_{ab}\nonumber\\
&&R_{\a\b\g\d}=g_{\a\g}g_{\b\d}-g_{\a\d}g_{\b\g};
\quad R_{\a\b}=4g_{\a\b}\nonumber\\
&&\bar{F}_{abcde}=\e_{abcde};\quad \bar{F}_{\a\b\g\d\e}=
\e_{\a\b\g\d\e},
\la{back}
\eea
where $a,b,c,\ldots $ and $\a ,\b ,\g ,\ldots$ are the AdS and 
the sphere indices respectively and $\eta_{ij}$ is the 
$4$-dimensional Minkowski metric.
We represent the gravitational field and the 4-form potential as
$$G_{MN}=g_{MN}+h_{MN};\quad A_{MNPQ}=\bar{A}_{MNPQ}+a_{MNPQ};
\quad F=\bar{F} +f.$$
Then the self-duality equation (\ref{ffe}) decomposed up to 
the second order looks as
\bea
f-f^*+T^{(1)}+T(h,f^*)+T(h)=0.
\la{ffeq}
\eea
Here we introduced the following notations 
\bea
 T_{M_1...M_5}^{(1)}&=&\frac{1}{2}h\bar{F}_{M_1...M_5}-
5h^K_{[M_1}\bar{F}_{M_2...M_4]K},\quad h=h_K^K
\nonumber   \\
T_{M_1...M_5}(h,f^*)&=&
\frac{1}{2}hf^*_{M_1...M_5}-5h^K_{[M_1}f^*_{M_2...M_4]K}
\nonumber  \\
\nonumber
T_{M_1...M_5}(h) &=&\frac{5}{2}hh^K_{[M_1}\bar{F}_{M_2...M_4]K}
-\left(\frac{1}{8}h^2+\frac{1}{4}h^{ML}h_{ML}\right)
\bar{F}_{M_1...M_5}\nonumber\\
&-&10h^{K_1}_{[M_1}h^{K_2}_{M_2}\bar{F}_{M_3M_4M_5]K_1K_2}. 
\la{deft}
\eea
Decomposing the Einstein equation (\ref{gre}) up to 
the second order, we get
\bea
&&R_{MN}^{(1)}+R_{MN}^{(2)}=
-\frac{4}{3!}h^{KL}\bar{F}_{MKM_1M_2M_3}\bar{F}_{NL}^{~~M_1M_2M_3}
\\
&&+\frac{1}{3!}(f_{MM_1...M_4}\bar{F}_{N}^{~M_1...M_4}+
\bar{F}_{MM_1...M_4}f_{N}^{~M_1...M_4})\nonumber\\
&&+\frac{4}{3!}h^{KL}h_L^S\bar{F}_{MKM_1M_2M_3}\bar{F}_{NS}^{~~M_1M_2M_3}
+\frac{2\cdot3}{3!}
h^{K_1S_1}h^{K_2S_2}\bar{F}_{MK_1K_2M_1M_2}\bar{F}_{NS_1S_2}^{~~~M_1M_2}
\nonumber \\
\nonumber
&&-\frac{4}{3!}h^{KS}(f_{MKM_1M_2M_3}\bar{F}_{NS}^{~~M_1M_2M_3}
+f_{NKM_1M_2M_3}\bar{F}_{MS}^{~~M_1M_2M_3})
+\frac{1}{3!}f_{MM_1...M_4}f_N^{~M_1...M_4}.
\la{greq}
\eea
Here
\bea
&&R_{MN}^{(1)}=\n_Kh_{MN}^K - \frac 12 \n_M\n_Nh_L^L \nonumber\\
&&R_{MN}^{(2)}=-\n_K(h^K_Lh_{MN}^L) + 
\frac 12 \n_N(h_{KL}\n_Mh^{KL})+
\frac 12 h_{MN}^K\n_Kh_L^L -h_{MK}^Lh_{NL}^K
\la{defr}
\eea
and we introduce a notation
\bea
h_{MN}^K=\frac{1}{2}(\n_Mh_N^K +\n_Nh_M^K -\n^Kh_{MN})
\la{defh}
\eea
In eqs.(\ref{ffeq}-\ref{defh}) and in what follows 
indices are raised and lowered 
by means of the background metric, and the covariant derivatives 
are with respect to the background metric, too.

The gauge symmetry of the equations of motion allows one to 
impose the de Donder gauge:
\bea
\n^\a h_{a\a }=\n^\a h_{(\a\b )}=\n^\a a_{M_1M_2M_3\a}=0;\quad  
h_{(\a\b )}\equiv h_{\a\b }-\frac{1}{5}g_{\a\b}h_\g^\g .
\la{ga}
\eea
This gauge choice does not remove all the gauge symmetry of the 
theory, for a detailed discussion of the residual symmetry 
see \cite{KRN}. As was shown in \cite{KRN}, the gauge 
condition (\ref{ga}) implies that the components of 
the 4-form potential of the form $a_{\a\b\g\d}$ and $a_{a\a\b\g}$ can be represented as follows:
\be
a_{\a\b\g\d}=\e_{\a\b\g\d\e}\n^\e b ;\quad
a_{a\a\b\g}=\e_{\a\b\g\d\e}\n^\d\phi_{a}^{\e}.
\la{phib}
\ee
It is also convenient to introduce the dual 1- and 2-forms for 
$a_{abcd}$ and $a_{abc\a}$:
\be
a_{abcd}=-\e_{abcde}Q^e;\quad
a_{abc\a}=-\e_{abcde}\phi_\a^{de}.
\la{aa}
\ee
Then the solution of the first-order self-duality equation
can be written as
\be
Q^a=\n^ab,\quad \phi_\a^{ab}=\n^{[a}\phi_\a^{b]}.
\la{qphi}
\ee
The quadratic action for physical fields of type IIB 
supergravity was found in \cite{AF3}. To write down 
the action we need to expand fields in spherical harmonics, 
and make some fields redefinition. We begin with the scalar
fields $b$ and $\pi \equiv h_\a^\a$. Expanding them into a 
set of scalar spherical harmonics\footnote{Here and in what 
follows we suppose that the spherical harmonics of all types
are orthonormal.}
\bea
\pi(x,y)=\sum\, \pi^{I_1}(x)Y^{I_1}(y);\quad 
b(x,y)=\sum\, b^{I_1}(x)Y^{I_1}(y); \quad \n_\b^2Y^k=-k(k+4)Y^k,
\nonumber
\eea
and making the fields redefinition \cite{LMRS}\footnote{We often denote
$\pi^{I_1}$ as $\pi_k$ and a similar notation for other fields.}
\bea
\pi_k=10ks_k+10(k+4)t_k;\quad b_k=-s_k+t_k
\la{redbpi}
\eea
we write the quadratic actions for the scalars $s^I$ and $t^I$ 
in the form
\bea
&&S(s)=\frac{4N^2}{(2\pi )^5}\int d^{5}x\sqrt{-g_a}\sum ~ 
\frac{32k(k-1)(k+2)}{k+1}\left( -\frac12 \n_as_k\n^as_k
-\frac12 k(k-4)s_k^2\right),
\la{as}\\
&&S(t)=\frac{4N^2}{(2\pi )^5}\int d^{5}x\sqrt{-g_a}\sum ~ 
\frac{32(k+2)(k+4)(k+5)}{k+3}
\left( -\frac12 \n_at_k\n^at_k
-\frac12 (k+4)(k+8)t_k^2\right).\nonumber\\ 
\la{at}
\eea
Now we expand the graviton on $AdS_5$ in scalar spherical harmonics
\bea
h_{ab}(x,y)=\sum\, h_{ab}^{I_1}(x)Y^{I_1}(y)
\nonumber
\eea
and  make the following shift of the gravitational fields:
\be
h_{ab}^k=\phi_{(ab)}^k +\n_{(a}\n_{b)}\zeta_k + 
\frac15 g_{ab}(\phi_{ck}^c -\frac35 \pi_k) ,
\la{redh}
\ee
where
\be
\zeta_k =\frac{4}{k+1}s_k +\frac{4}{k+3}t_k .
\la{zeta}
\ee
Then the zero mode $\phi_{ab}^0\equiv \phi_{ab}$ describes a 
graviton on $AdS_5$ 
with the standard action
\bea
S(\phi_{ab})=&&\frac{4N^2}{(2\pi )^5}\int d^{5}x\sqrt{-g_a}\left( -\frac{1}{4}\n_c\phi_{ab}\n^c\phi^{ab}+
\frac{1}{2}\n_a\phi^{ab}\n^c\phi_{cb}-
\frac{1}{2}\n_a\phi_c^c\n_b\phi^{ba}
\right.\nonumber\\
&&+\frac{1}{4}\n_c\phi_a^a\n^c\phi^{b}_b
+\left.\frac{1}{2}\phi_{ab}\phi^{ab}+
\frac{1}{2}(\phi_{a}^{a})^2\right)
\la{agr0}
\eea
and the action for the traceless symmetric tensor 
fields $\phi_{ab}^k$ has the form
\bea
S(\phi_{(ab)}^k)=&&\frac{4N^2}{(2\pi )^5}\int d^{5}x\sqrt{-g_a}\sum ~\left( -\frac{1}{4}\n_c\phi_{(ab)}^k\n^c\phi^{(ab)}_k+
\frac{1}{2}\n_a\phi^{(ab)}_k\n^c\phi_{(cb)}^k\right.\nonumber\\&&\left. -
\frac{1}{4}(k^2+4k-2)\phi_{(ab)}^k\phi^{(ab)}_k\right)
\la{agr2}
\eea
As was shown in \cite{KRN} the fields $\phi_{ck}^c$ are nondynamical 
and vanish on shell at the linearized level.

Expanding vector fields $h_{a\a}$ and $\phi_{a\a}$  into a 
set of vector spherical harmonics
\bea
&&h_{a\a}(x,y)=\sum\, hh_a^{I_5}(x)Y_\a^{I_5}(y);\quad 
\phi_{a\a}(x,y)=\sum\, \phi_a^{I_5}(x)Y_\a^{I_5}(y); \nonumber\\
&&(\n_\b ^2-4)Y_\a^k=-(k+1)(k+3)Y_\a^k,
\nonumber
\eea
and making the change of variables \cite{KRN}
\bea
A_a^k=h_a^k -4(k+3)\phi_a^k;\quad C_a^k=h_a^k +4(k+1)\phi_a^k
\la{redhphi}
\eea
we present the actions for the vector fields in the form
\bea
&&S(A)=\frac{4N^2}{(2\pi )^5}\int d^{5}x\sqrt{-g_a}\sum  
\frac{k+1}{2(k+2)}\left(-\frac14 (F_{ab}(A^k))^2
-\frac12 (k^2-1)(A_a^k)^2\right)
\la{aA}\\
&&S(C)=\frac{4N^2}{(2\pi )^5}\int d^{5}x\sqrt{-g_a}\sum  
\frac{k+3}{2(k+2)}\left(-\frac14 (F_{ab}(C^k))^2
-\frac12 (k+3)(k+5)(C_a^k)^2\right) 
\la{aC}
\eea
where $F_{ab}(A)=\p_aA_b-\p_bA_a$. 
Finally, expanding the graviton on the sphere in tensor harmonics
\bea
h_{(\a\b)}(x,y)=\sum\, \phi^{I_{14}}(x)Y_{(\a\b)}^{I_{14}}(y);\quad 
(\n_\g^2-10)Y_{(\a\b)}^k=-(k^2+4k+8)Y_{(\a\b)}^k,
\nonumber
\eea
we write the action for the scalars $\phi_k$ in the form
\bea
S(\phi)=\frac{4N^2}{(2\pi )^5}\int d^{5}x\sqrt{-g_a}\sum ~ 
\left( -\frac14 \n_a\phi_k\n^a\phi_k -\frac14 k(k+4)\phi_k^2\right)
\la{aphi}
\eea
\section{Cubic couplings of scalars}
The aim of this section is to find the cubic couplings
of the scalar fields $t_k$ and $\phi_k$ with a pair of scalars $s_k$.
This can be achieved by finding the quadratic contribution of the scalars 
$s_k$ to the equations of motion  for $t_k$ and $\phi_k$
respectively with a subsequent
reconstruction of the corresponding Lagrangian vertex. 

\vskip 0.4cm
{\it 4.1 Cubic Couplings of $t_k$}
\vskip 0.4cm
\noindent
Since $t_k$ appear as the mixture of fields $\pi_k$ and $b_k$
we begin by considering the equations of motion for these fields. 
Restricting in (\ref{greq}) indices $M$ and $N$ to the sphere
and taking into account the gauge conditions (\ref{ga}), (\ref{phib})   
we find that Einstein equation (\ref{greq}) results in
\bea
\la{bG}
&&\frac{1}{10}g_{\a\b}\l (\n_M\n^M-32)\pi+80\n_{\g}\n^{\g}b\r
+\frac{1}{2}\n_{\a}\n_{\b}\phi_a^a=\\
\nonumber
&+&\frac{1}{10}g_{\a\b}\n_a(h^{ab}\n_b \pi)+\frac{3}{100}\n_{\a}\pi\n_{\b}\pi
+\frac{3}{50}\pi\n_{\a}\n_{\b}\pi+\frac{1}{4}\n_{\a}h_{ab}\n_{\b}h^{ab}
+\frac{1}{2}h_{ab}\n_{\a}\n_{\b}h^{ab}\\
\nonumber
&+&8\n_{\a}\n^a b\n_{\b}\n_a b 
-4g_{\a\b}\left( 
\n_{\g}\n^a b \n^{\g}\n_a b+\n_{\g}^2b \n_{\d}^2b +\frac{2}{5}\pi^2
-\frac{8}{5}\pi\n_{\g}^2b-\frac{1}{200}\n_{\g}(\pi\n^{\g}\pi)  
\right),
\eea
where $\phi_a^a= h_a^a+\frac{3}{5}\pi$
in accordance with (\ref{redh}). 
Note that we have omitted all the linear terms that are projected
out under the projection onto the spherical harmonics 
$\n_{(\a}\n_{\b)}Y^I$
or $Y^I$ and accounted only for the quadratic terms that contain 
after the field redefinition (\ref{redbpi}) and (\ref{redh}) 
two scalars $s_k$. In particular 
the scalars $s_k$ appear after redefinition (\ref{redh})
for the gravitational field $h_{ab}$. 

Equation (\ref{bG}) implies then the following two equations
\bea
\la{ctr}
\n_{(\a}\n_{\b)}\phi_a^a &=&
\frac{3}{50}\n_{(\a}\pi\n_{\b)}\pi
+\frac{3}{25}\pi\n_{(\a}\n_{\b)}\pi \\
\nonumber
&+&\frac{1}{2}\n_{(\a}h_{ab}\n_{\b)}h^{ab}
+h_{ab}\n_{(\a}\n_{\b)}h^{ab}+16\n_{(\a}\n^a b\n_{\b)}\n_a b 
\eea
and
\bea
\la{1bp}
&& (\n_M\n^M-32)\pi+80\n_{\g}\n^{\g}b
+\n_{\a}\n^{\a}\phi_a^a=\\
\nonumber
&&\n_a(h^{ab}\n_b \pi)+\frac{13}{50}\n_{\a}\pi\n^{\a}\pi
+\frac{8}{25}\pi\n_{\a}\n^{\a}\pi+\frac{1}{2}\n_{\a}h_{ab}\n^{\a}h^{ab}
+h_{ab}\n_{\a}\n^{\a}h^{ab}\\
\nonumber
&&-24\n_{\a}\n^a b\n^{\a}\n_a b 
-40\n_{\g}^2b \n_{\d}^2b -16\pi^2
+64\pi\n_{\g}^2b    
\eea
that are obtained by decoupling from (\ref{bG}) the trace part.
Projecting eq.(\ref{ctr}) onto $\n_{\a}\n_{\b}Y^I$ one can solve it
for $\phi_a^a$ and substituting the result in (\ref{1bp})
obtain the close equation for $\pi$ and $b$.

According to \cite{KRN} the second equation involving
the fields $\pi$ and $b$ is found by considering
the component of the self-duality equation (\ref{ffeq})
involving one sphere and four AdS indices, and the component 
with five AdS indices.  In our case these components 
read as  
\bea
\la{sd1}
\n_{\a}\l a_{a_1...a_4}+\eps_{a_1...a_5}\n^{a_5}b \r =
\eps_{a_1...a_4 a}
\l \frac{3}{5}\pi\n^a\n_{\a}b+h^{ab}\n_b\n_{\a}b \r
\eea
and
\bea
\la{sd2}
5\n_{[a_1}a_{a_2...a_5]}=
\eps_{a_1...a_5}\l \n_{\g}^2 b+\frac{1}{2}\phi_a^a-\frac{4}{5}\pi
-\frac{4}{5}\pi\n_{\g}^2 b-\frac{1}{4}h_{ab}h^{ab}+\frac{37}{100}\pi^2 \r .
\eea
Projecting (\ref{sd1}) onto $\n_{\a}Y^I$ one finds $a_{a_1...a_5}$.
Substituting then $a_{a_1...a_5}$ as well as previously found 
$\phi_a^a$ into (\ref{sd2}) one obtains the  equation 
for $\pi$ and $b$.  

The required equation for $t_k$ is then obtained by substituting 
the redefinition (\ref{redbpi}) in (\ref{1bp}-\ref{sd2}) and by
eliminating all the terms linear 
in $s_k$. Skipping all the computational details we write
down the equation for $t^{I}$ that is found to be of the form
\bea
\nonumber
(\n_a\n^a-(k_3+4)(k_3+8))t^{I_3}=D_{123}s^{I_1}s^{I_2}
+E_{123}\n^a s^{I_1}\n_a s^{I_2}
+F_{123}\n_{(a}\n_{b)} s^{I_1}\n^{(a}\n^{b)} s^{I_2}.
\eea
To remove the derivative terms we perform
the appropriate redefinition of $t^I$ similar to (\ref{red}):
\bea
\nonumber
t^{I_3}= t'^{I_3}+\sum_{I_1,I_2}\left( 
J_{I_1I_2I_3}s'^{I_1}s'^{I_2}+L_{I_1I_2I_3}\n^a s'^{I_1}\n_a s'^{I_2}
\right).
\eea
Introducing the notation $a_{123}=\int Y^{I_1} Y^{I_2} Y^{I_3}$
we quote the final answer
\bea
\nonumber
&&(\n_a\n^a-(k_3+4)(k_3+8))t^{I_3}=-t_{I_1I_2I_3}s^{I_2}s^{I_3},\\
\nonumber
&&t_{I_1I_2I_3}=a_{123}\frac{4(\S+4)(\a_1+2)(\a_2+2)\a_3
(\a_3-1)(\a_3-2)(\a_3-3)(\a_3-4)}
{(k_1+1)(k_2+1)(k_3+2)(k_3+4)(k_3+5)},
\eea 
where $\a_3=\frac{1}{2}(k_1+k_2-k_3)$, $\S=k_1+k_2+k_3$. 

Taking into account the normalization of the quadratic
action for $t_k$ fields (\ref{at}) we obtain the corresponding
vertex
\bea
\nonumber
S_{tss}=\frac{4N^2}{(2\pi )^5} T_{I_1I_2I_3}
\int \sqrt{-g_a}~s^{I_1}s^{I_2}t^{I_3}
\eea
with 
\bea
\la{vtss}
T_{I_1I_2I_3}=
a_{123}\frac{2^7(\S+4)(\a_1+2)(\a_2+2)\a_3(\a_3-1)(\a_3-2)(\a_3-3)(\a_3-4)}
{(k_1+1)(k_2+1)(k_3+3)}.
\eea
\vskip 0.4cm
{\it 4.2 Cubic Couplings of $\phi_k$}
\vskip 0.4cm
\noindent
To find equations of motion for the fields $\phi_k$ coming from  
the graviton on the sphere we  again consider
 eq.(\ref{greq}) for the indices $M=\a$, $N=\b$:
\bea
\nonumber
(\n_M\n^M-2)h_{(\a\b)}&=&\frac{3}{50}\n_{(\a}\pi\n_{\b)}\pi
+\frac{3}{25}\pi\n_{(\a}\n_{\b)}\pi+
\frac{1}{2}\n_{(\a}h_{ab}\n_{\b)}h^{ab}\\
\nonumber
&+&h_{ab}\n_{(\a}\n_{\b)}h^{ab}+16\n_{(\a}\n^a b\n_{\b)}\n_a b, 
\eea
where this time all the linear terms that are projected out 
under the projection on $Y_{(\a\b)}$ were omitted.

Introducing the notation 
$p_{123}=\int \n^{\a}Y^{I_1}\n^{\b}Y^{I_2}Y_{(\a\b)}^{I_3}$
and projecting both sides of the last equation on  $Y_{(\a\b)}$ 
we get an equation for $\phi$:
\bea
\nonumber
(\n_a\n^a-k_3(k_3+4))\phi^{I_3}=
p_{123}\l -\frac{3}{50}\pi^{I_1}\pi^{I_2}
-\frac{1}{2}h_{ab}^{I_1}h^{ab}_{I_2}
+16\n^a b^{I_1}\n_a b^{I_2} \r .
\eea

Finally leaving on the r.h.s. only the contribution of the 
scalars $s_k$ we obtain
\bea
\nonumber
&&(\n_a\n^a-k_3(k_3+4))\phi^{I_3}=-\frac{p_{123}}
{5(k_1+1)(k_2+1)}\times \\
\nonumber
&&
\l
48k_1k_2(k_1+1)(k_2+1)s^{I_1}s^{I_2}-80(k_1+1)(k_2+1)\n_a s^{I_1}\n^a s^{I_2}
+40\n_{(a}\n_{b)}s^{I_1}\n^{(a}\n^{b)}s^{I_2}
\r .
\eea
Performing again a shift of $\phi^I$ to get rid of the 
derivative terms one arrives at 
\bea
\nonumber
(\n_a\n^a-k_3(k_3+4))\phi^{I_3}=-\frac{8 p_{123}\S (\S+2)}
{(k_1+1)(k_2+1)}(\a_3-1)(\a_3-2).
\eea
Taking into account the normalization of the quadratic action
for $\phi_k$ we can read off the corresponding vertex
$S_{ss\phi}$:
\bea
S_{ss\phi}=\frac{4N^2}{(2\pi )^5}\Phi_{I_1I_2I_3}
\int \sqrt{-g_a}~s^{I_1}s^{I_2}\phi^{I_3},
\la{ssphi}
\eea
where 
$$
\Phi_{I_1I_2I_3}=\frac{4 p_{123}\S (\S+2)}
{(k_1+1)(k_2+1)}(\a_3-1)(\a_3-2).
$$
\vskip 0.4cm
{\it 4.3 Three-point Functions}
\vskip 0.4cm
\noindent
Recall that two- and three-point correlation functions of operators 
${\cal O}_{\D}$ in a boundary conformal field theory corresponding  
to scalar fields on AdS are given by \cite{FMMR}:
\bea
\langle {\cal O}_{\D}(\x ){\cal O}_{\D}(\y )\rangle =\frac{2}{\pi^2}
\frac{\theta (\D-1)(\D-2)^2}{|\x-\y|^{2\D}},
\eea
\bea
\langle {\cal O}_{\D_1}(\x ){\cal O}_{\D_2}(\y ){\cal O}_{\D_3}(\z )\rangle
=\frac{\lambda_{123}}
{|\x-\y|^{\D_1+\D_2-\D_3}|\x-\z|^{\D_1+\D_3-\D_2}|\y-\z|^{\D_3+\D_2-\D_1}},
\eea
where $\lambda_{123}$ is given by 
$$
\lambda_{123}=-\varphi_{123}
\frac{ \G[\frac12 \l \D_1+\D_2+\D_3-4 \r]
\G[\bar{\D}_1 ]
\G[\bar{\D}_2 ]
\G[\bar{\D}_3 ]
}{2\pi^4\G(\D_1-2)\G(\D_2-2)\G(\D_3-2)}
$$
and $\bar{\D}_1=\frac{1}{2}(\D_2+\D_3-\D_1)$.
Here $\varphi_{123}$ stands for the coupling of scalar fields
(that is a doubled interaction vertex for the fields we consider)
and $\theta$ denotes the normalization constant of their 
quadratic action. Taking into account that 
a scalar $t^{I_3}$ $(\phi^{I_3})$ corresponds to a YM operator 
${\cal O}_{\D_3}$
with the conformal weight $\D_3=k_3+8$ $(\D_3=k_3+4)$, we,
therefore find correlation functions of two extended CPOs
with this operator. The constant $\lambda_{123}$
reads for both cases as follows:
\bea
\nonumber
\lambda_{123}(t)=-\frac{4N^2}{(2\pi )^5}\frac{2^8}{\pi^4}
\frac{\G\l \frac{1}{2}\S+3 \r
\G(\a_1+4)\G(\a_2+4)\G(\a_3+1)(\a_1+2)(\a_2+2)
}{(k_1+1)(k_2+1)(k_3+3)\G(k_1-2)\G(k_2-2)\G(k_3+6)}a_{123}
\eea
and 
\bea
\nonumber
\lambda_{123}(\phi)=-\frac{4N^2}{(2\pi )^5}\frac{2^4}{\pi^4}
\frac{\G\l \frac{1}{2}\S+2 \r
\G(\a_1+2)\G(\a_2+2)\G(\a_3)
}{(k_1+1)(k_2+1)
\G(k_1-2)\G(k_2-2)\G(k_3+2)}p_{123}.
\eea
Taking into account the normalization of the two-point
functions one can introduce the normalized extended CPO \cite{LMRS}:
\bea
\la{nCPO}
O_{\D}=\frac{(2\pi )^{5/2}}{2N}\frac{\pi}{8(k-1)(k-2)}\l \frac{k+1}{k(k+2)}\r^{1/2}{\cal O}_{\D}
\eea
as well as the normalized gauge theory operator corresponding
to scalar $t_k$:
$$
O_{\D}=\frac{(2\pi )^{5/2}}{2N}\frac{\pi}{8(k+6)}\l \frac{k+3}{(k+2)(k+4)(k+5)(k+7)}\r^{1/2}
{\cal O}_{\D}, ~~~\D = k+8
$$
and to scalar $\phi_k$:
$$
O_{\D}=\frac{(2\pi )^{5/2}}{2N} \frac{\pi}{(k+3)^{1/2}(k+2)}
{\cal O}_{\D}, ~~~\D = k+4.
$$
With these formulae at hand we can finally write down 
the normalized constants:
\bea
\nonumber
\lambda^{norm}_{123}(t)&=&-\frac{(2\pi )^{5/2}}{N}\frac{1}{(2\pi)^{5/2}}
\l
\frac{k_1k_2(k_3+1)(k_3+7)}{(k_3+3)(k_3+4)(k_3+5)}
\r^{1/2} \\
\nonumber
&\times &  
\frac{\G(\a_1+4)(\a_1+2)}{\a_1!}\frac{\G(\a_2+4)(\a_2+2)}{\a_2!}
\frac{k_3!}{\G(k_3+8)}\langle {\cal C}^{I_1}{\cal C}^{I_2}{\cal C}^{I_3}\rangle
\eea
and 
\bea
\nonumber
\lambda^{norm}_{123}(\phi)=-\frac{(2\pi )^{5/2}}{N}
\frac{(\a_1+1)(\a_2+1)}{4(2\pi)^{5/2}}
\l
\frac{k_1k_2}{(k_3+1)(k_3+2)(k_3+3)}
\r^{1/2} P_{123} .
\eea
Here we used explicit expressions for $a_{123}$ and $p_{123}$ 
from the Appendix.
\section{Cubic couplings of second rank tensors  with $s^{I}$}
\vskip 0.4cm
{\it 5.1 Cubic Couplings}
\vskip 0.4cm
\noindent
Clearly the coupling of the symmetric second rank tensor
$\phi_{(ab)}^k$ with a pair of scalars $s_k$ can be found 
by studing the corrected equation of motion for $\phi_{(ab)}^k$.
The most simple way consists however in finding the 
equations of motion for the field $s_k$ corrected 
by the quadratic terms each containing one field 
$\phi_{(ab)}^k$ and $s_k$. This is explained by noting
that the field $\phi_{(ab)}^k$ is transverse on-shell and
therefore the interaction term, being in the latter case a Lorentz scalar 
does not contain derivatives acting on  $\phi_{(ab)}^k$.
As a consequence the additional shift needed to get rid 
of derivative terms is not required. 

Since the field $s_k$ appear 
as the mixture (\ref{redbpi}) of $\pi$ and $b$, the equation
for $s_k$ again follows from the system (\ref{ctr})-(\ref{sd2}).
Clearly this time eqs.(\ref{ctr}) and (\ref{1bp}) read as
\bea
\la{ctr1}
\n_{(\a}\n_{\b)}\phi_a^a =
\n_{(\a}\phi_{(ab)}\n_{\b)} \n^a\n^b \zeta
+\phi_{(ab)}\n_{(\a}\n_{\b)}\n^a\n^b\zeta
+\n_a\n_b\zeta \n_{(\a}\n_{\b)}\phi^{(ab)}
\eea
and
\bea
\la{2bp}
&& (\n_M\n^M-32)\pi+80\n_{\g}\n^{\g}b
+\n_{\a}\n^{\a}\phi_a^a=\\
\nonumber
&&\n_a(\phi^{(ab)}\n_b \pi)+
\n_{(\a}\phi_{(ab)}\n_{\b)}\n^a\n^b \zeta
+\phi_{(ab)}\n_{(\a}\n_{\b)}\n^a\n^b\zeta
+\n_a\n_b\zeta \n_{(\a}\n_{\b)}\phi^{(ab)}  
\eea
where we have used representation (\ref{redh}) for the graviton 
field $h_{ab}$ and left only the terms contributing to the vertex
under consideration. By this reason the coefficients $\zeta_k$ in
$\zeta=\int \zeta^I Y^I$ are reduced now to $\zeta_k=\frac{4}{k+1}s_k$
in comparison with (\ref{zeta}).

Again projecting eq.(\ref{ctr1}) onto $\n_{\a}\n_{\b}Y^I$ one solves
for $\phi_a^a$ and after substitution of the solution into (\ref{2bp})
one obtains a closed form equation for $\pi$ and $b$.

The second equation for  $\pi$ and $b$ follows from eqs.(\ref{sd1})
and (\ref{sd2}) that now acquire the form  
\bea
\la{sd1'}
\n_{\a}\l a_{a_1...a_4}+\eps_{a_1...a_5}\n^{a_5}b \r =
\eps_{a_1...a_4 a}
\l \phi^{(ab)}\n_b\n_{\a}b \r
\eea
and
\bea
\la{sd2'}
5\n_{[a_1}a_{a_2...a_5]}=
\eps_{a_1...a_5}\l \n_{\g}^2 b+\frac{1}{2}\phi_a^a-\frac{4}{5}\pi
-\frac{1}{2}\phi_{(ab)}\n^a\n^b\zeta \r .
\eea

Omitting the straightforward but lengthy algebraic manipulations we
write down the final answer for the Lagrangian vertex
describing the interaction of the symmetric second rank tensor $\phi_{(ab)}$
with scalars $s^I$:
\bea
\nonumber
S_{ssg}=\frac{4N^2}{(2\pi )^5}G_{I_1I_2I_3}\int \sqrt {-g_a} 
\n^a s^{I_1}\n^b s^{I_2}\phi_{(ab)}^{I_3},
\eea
where $G_{I_1I_2I_3}$ is found to be 
\bea
\nonumber
G_{I_1I_2I_3}=\frac{4(\S+2)(\S+4)\a_3(\a_3-1)}{(k_1+1)(k_2+1)} a_{123}.
\eea
\vskip 0.4cm
{\it 5.2 Three-point Functions}
\vskip 0.4cm
\noindent
Denote by ${\cal T}_{ij}^I$ the operator in SYM of the conformal
weight $\D_G=k+4$ that corresponds to the AdS field $\phi_{(ab)}$. 
To compute the three-point correlation function of this 
operator with extended CPOs in the boundary conformal field theory 
one needs the 
bulk-to-boundary propagator for the field $ \phi_{(ab)}^{I}$.
In principle this can be extracted from the momentum space results 
of \cite{P}. In the case of three-point correlators it is 
however more convenient to deal directly with the $x$-space 
propagator.  

Recall that the linearized equations of motion for 
$ \phi_{(ab)}^{I}$ read as
\bea
\la{gre1}
\n_{c}\n^{c}\phi_{(ab)}^I+(2-k^2-4k)\phi_{(ab)}^I=0,\quad 
\n^{b}\phi_{(a b)}^I =0.
\eea
Now one can easily check that the following function
\bea
\la{grg}
G_{ab~ij}(\o_0,\x)=\frac{\D_G+1}{\D_G-1}
\o_0^2{\cal K}_{\D_G}(\o,\x )J_{ak}(\o-\x)J_{bl}(\o-\x)
{\cal E}_{ij, kl}
\eea
is the bulk-to-boundary Green function for eq.(\ref{gre1}). Here
${\cal E}_{ij, kl}$ denotes the traceless symmetric
projector:
$$
{\cal E}_{ij, kl}=\frac{1}{2}(\d_{ik}\d_{jl}+\d_{il}\d_{kj})-\frac{1}{4}
\d_{ij}\d_{kl},
$$ 
${\cal K}_{\D} (\o,\x)$ is a bulk-to-boundary propagator 
for a scalar field corresponding to an operator of conformal dimension $\D$:
\bea
\la{skprop}
{\cal K}_{\D} (\o,\x)=
c_{\D}\frac{\o_0^{\D}}{(\o_0^2+({\vec{\o}}-\x)^2)^\D},\quad
c_{\D}=\frac{\G(\D)}{\pi^2 \G(\D-2)},
\eea
and $J_{ab}(x)=\d_{ab}-2\frac{x_a x_b}{x^2}$.

Note that function (\ref{grg}) satisfies the transversality
condition $\n^{a}G_{ab~ij}=0$. 
The normalization
constant $\frac{\D_G+1}{\D_G-1}$ in (\ref{grg})
is fixed by requiring the corresponding solution 
of (\ref{gre1}) to reproduce correctly the boundary data in the 
limit $\o_0\to 0$. In the case of vanishing AdS mass eq.(\ref{grg})
turns into the graviton bulk-to-boundary propagator \cite{LT}.

Having discussed the propagator for $\phi_{(ab)}$ we come
back to the three-point correlator that now reads as  
\bea
\la{OOT}
\langle {\cal O}^{I_1}(\x){\cal O}^{I_2}(\y){\cal T}_{ij}^{I_3}(\z)\rangle 
=-\frac{8N^2}{(2\pi )^5}G_{I_1I_2I_3}\int 
\frac{d^5\o}{\o_0^5}\o_0^4\n_a\n_b
{\cal K}_{\D_1}(\o,\x){\cal K}_{\D_2}(\o,\y)G_{ab~ij}^{I_3}(\o, \z).
\eea 
By the conformal symmetry this correlator is defined up
to the normalization constant $\b_{123}$:
\bea 
\nonumber
&&\langle {\cal O}^{I_1}(\x){\cal O}^{I_2}(\y){\cal T}_{ij}^{I_3}(\z)\rangle 
=\\
&&\frac{\b_{123}}
{|\x-\y|^{\D_1+\D_2-\D_G}|\x-\z|^{\D_1+\D_G-\D_2} |\y-\z|^{\D_2+\D_G-\D_1}}
\l \frac{Z_iZ_j}{Z^2}-\frac{1}{d}\d_{ij}\r ,
\nonumber
\eea 
where
\bea
\la{defzi}
Z_i=\frac{(\x-\z)_i}{(\x-\z)^2}-\frac{(\y-\z)_i}{(\y-\z)^2}.
\eea
This constant is then found by explicit evaluation of 
integral (\ref{OOT}):
\bea
\beta_{123}&=&-\frac{4N^2}{(2\pi )^5}
4\pi^2c_{\D_1}c_{\D_2}c_{\D_G} G_{I_1I_2I_3}
\frac{\D_G+1}{\D_G-1}
\frac{\G\l \frac{1}{2} (\D_1+\D_2+\D_G-2) \r }{\G(\D_G+2) } \\
\nonumber
& \times &
\frac{\G\l\frac{1}{2} (\D_1+\D_G-\D_2+2) \r
\G\l\frac{1}{2} (\D_2+\D_G-\D_1+2) \r
\G\l\frac{1}{2} (\D_1+\D_2-\D_G+2) \r  }
{ \G(\D_1)\G(\D_2) }
\eea
Substituting here the normalization constants and $G_{I_1I_2I_3}$
we finally find
\bea
\nonumber
\beta_{123}&=&-\frac{4N^2}{(2\pi )^5}\frac{64}{\pi^4}\l \frac{k_3+2}{(k_1+1)(k_2+1)}\r
\frac{\G\l \frac{1}{2}\S+3 \r 
\G\l\a_1+3 \r
\G\l\a_2+3\r
\G\l\a_3+1\r  }
{\G(k_1-2)\G(k_2-2)\G(k_3+5) }a_{123}
\eea

The two-point correlation function of the YM operator ${\cal T}_{ij}$
corresponding to the symmetric second rank tensor field $\phi_{(ab)}$
was computed in \cite{P}
$$
\langle {\cal T}_{ij}^I(\x){\cal T}_{kl}^J(\y)\rangle =\frac{4N^2}{(2\pi )^5}
\frac{1}{\pi^2}(\D_G-2)^2(\D_G+1)\frac{\d^{IJ}}{|\x-\y|^{2\D_G}}
{\cal E}_{ij~i'j'}J_{i'k}(\x-\y)J_{j'l}(\x-\y).
$$ 
Therefore, introducing the normalized operator 
$$
T_{ij}^I=\frac{(2\pi )^{5/2}}{2N}
\frac{\pi}{(\D_G-2)(\D_G+1)^{1/2}}{\cal T}_{ij}^I 
$$
one obtains the correlation function of two normalized CPO's and $T_{ij}^I$
with the constant $\b_{123}^{norm}$:
\bea
\nonumber
\beta_{123}^{norm}&=&-\frac{(2\pi )^{5/2}}{N}
\frac{1}{2^{3/2}\pi^{5/2}}
\l k_1 k_2(k_3+1)(k_3+2)(k_3+5)\r^{1/2}
\\
\nonumber
&\times & 
\frac{(\a_1+1)(\a_1+2)(\a_2+1)(\a_2+2)}
{(k_3+1)(k_3+2)(k_3+3)(k_3+4)(k_3+5)} 
\langle {\cal C}^{I_1}{\cal C}^{I_2}{\cal C}^{I_3}\rangle ,
\eea
where the explicit expression for $a_{123}$ was used. Note that 
the variable $\a_3$ completely dissappeared from the final answer.
\section{Cubic couplings of two scalars $s^I$ with vector
fields}
\vskip 0.4cm
{\it 6.1 Cubic Couplings}
\vskip 0.4cm
\noindent
To obtain cubic couplings of two scalars $s^I$ with vectors
fields we need equations of motion for the vector fields up to
the second order. The equations
of motion for the vector fields $\phi_a^\a$ can be derived 
from the following components of the self-duality equation
\bea
&&f_{\a abcd}-f^*_{\a abcd}+T^{(1)}_{\a abcd}+
T(h,f^*)_{\a abcd}+T(h)_{\a abcd}=0,
\la{phi1}\\
&&f_{\a\b abc}-f^*_{\a\b abc}+T^{(1)}_{\a\b abc}+
T(h,f^*)_{\a\b abc}+T(h)_{\a\b abc}=0.
\la{phi2}
\eea
From the definition of $f$ we have
$$
f_{\a\b abc}= 2\n_{[\a}a_{\b ]abc},\quad
f^*_{\a\b abc}= \e_{abcde}(\n^d\n_\a\phi^e_\b -
\n^d\n_\b\phi^e_\a ).
$$
Here we omitted  all terms dependent on the components of the 
4-form potential of the form $a_{ab\a\b}$  which are not 
relevant for the cubic
couplings under consideration. From the definition of
the tensors $T$ (\ref{deft}) we can easily see that
$$T^{(1)}_{\a\b abc}=
T(h,f^*)_{\a\b abc}= T(h)_{\a\b abc}=0,
$$
if we keep only terms which may give a contribution to 
the cubic couplings.
Thus eq.(\ref{phi2}) does not get relevant quadratic 
corrections, and, therefore, 
\bea
a_{\a abc}=\e_{abcde}\n^{d}\phi_\a^{e}
\la{phi3}
\eea
Taking into account eq.(\ref{phi3}) and formulas (\ref{deft}) for 
the tensors $T$, one can rewrite eq.(\ref{phi1}) in the form
\bea
(\n_b^2 +\n_\b^2 -4)\phi_\a^a -\n_b\n^a\phi_\a^b -h_\a^a +
\frac12 h_b^b\n^a\n_\a b-\frac{3}{10}\pi\n^a\n_\a b-
h^{ab}\n_b\n_\a b=0
\la{phi4}
\eea
Here we have omitted all terms that are projected out under 
the projection onto $Y_\a$. 
Expanding all the fields in spherical harmonics and using 
eqs.(\ref{redbpi}-\ref{zeta}), we obtain equations of motion
for the vector fields $\phi_a^{I}$
\bea
&&\n_b^2\phi_a^3 -\n^b\n_a\phi_b^3 -(k_3+1)(k_3+3)\phi_a^3
-h_a^3 =\nonumber\\
&&-t_{123}\left( \frac{4k_2(k_2+2)}{k_2+1}s_2\n_as_1 +
\frac{4}{k_2+1}\n_a\n_bs_2\n^bs_1\right),
\la{phi5}
\eea
where $t_{123}\equiv t_{I_1I_2I_3}=\int\n^\a Y^{I_1}Y^{I_2}Y_\a^{I_3}$, 
$\phi_3$ means $\phi^{I_3}$ and so on, and summation over 1 and 2 
is assumed.   

Now we proceed with the equations of motion for $h^\a_a$. 
These equations can be derived from the $a,\a$ components of 
eq.(\ref{greq}). Omitting all intermediate calculations, 
we present the equations in the form
\bea
&&\n_b^2h_a^3 -\n^b\n_ah_b^3 -((k_3+1)(k_3+3)+8)h_a^3
-16(k_3+1)(k_3+3)\phi_a^3 =\nonumber\\
&&2t_{123}f(k_1,k_2)s_1\n_as_2 -
16t_{123}\frac{k_2-5}{k_1+1}\n_a\n_bs_1\n^bs_2
+\frac{8t_{123}}{(k_1+1)(k_2+1)}\n_a\n_b\n_cs_2\n^b\n^cs_1,\nonumber\\
\la{phi6}
\eea
where
\bea
&&f(k_1,k_2)=\frac{2k_1k_2(k_2-1)}{k_2+1}-
 \frac{4k_1(k_2^2-4k_2-4)}{k_2+1}-
\frac{5k_1(k_1-1)k_2(k_2-1)}{(k_1+1)(k_2+1)}+\nonumber\\
&&\frac{4k_1(k_1-4)k_2(k_2-1)}{(k_1+1)(k_2+1)}-
\frac{8k_1(k_1-4)}{(k_1+1)(k_2+1)}-k_1k_2+
48k_1-8k_1(k_1+4). 
\nonumber
\eea
The equations of motion for vector fields $A$ and $C$ are linear 
combinations of the two above and can be written in the form
\bea
&&\n_b^2V_a^3 -\n^b\n_aV_b^3 -m_3^2V_a^3=\n_aV^3 +
D_{123}s_1\n_as_2+\nonumber\\
&&\quad E_{123}\n^bs_1\n_a\n_bs_2+
F_{123}\n^b\n^cs_1\n_a\n_b\n_cs_2,
\la{phi7}
\eea
where $V$ may be either $A$ or $C$, and the constants 
$D$, $E$, $F$ are antisymmetric with respect to the permutation
of the indices 1 and 2.
We can remove the higher-derivative terms from the equation by 
means of the following field redefinition
\bea
V_a^3\to V_a^3 -\frac{1}{m_3^2}\n_a\tilde{V}^3 +
J_{123}s_1\n_as_2 +
L_{123}\n^bs_1\n_a\n_bs_2 ,
\la{phi8}
\eea
where
\bea
&&2L_{123}=F_{123}\nonumber\\
&&2J_{123}+L_{123}(m_1^2+m_2^2-m_3^2-12)=E_{123}\nonumber\\
&&\tilde{V}^3=V^3-(J_{123}-2L_{123})m_1^2s_1s_2-
L_{123}m_1^2\n_bs_1\n^bs_2
\nonumber
\eea
Then eq.(\ref{phi7}) acquires the form
\bea
\n_b^2V_a^{I_3} -\n^b\n_aV_b^{I_3} -m_3^2V_a^{I_3} +
\sum_{I_1,I_2}v_{I_1I_2I_3}s^{I_1}\n_as^{I_2}=0,
\la{phi9}
\eea
where
\bea
v_{I_1I_2I_3}=-D_{I_1I_2I_3}+
J_{I_1I_2I_3}(m_1^2+m_2^2-m_3^2)
-  2L_{I_1I_2I_3}(m_1^2+m_2^2)
\la{phi10}
\eea
A straightforward calculation of the constants $v$ gives
\bea
\la{phi11}
&&v_{I_1I_2I_3}(A)=\frac{4(\a_3-1/2)(\S-1)(\S+1)(\S+3)}
{(k_1+1)(k_2+1)}t_{123}\\
&&v_{I_1I_2I_3}(C)=\frac{16(\a_3-1/2)(\a_3-3/2)(\a_3-5/2)(\S+3)}
{(k_1+1)(k_2+1)}t_{123}
\la{phi12}
\eea
Taking into account the normalization of the quadratic actions (\ref{aA})
and (\ref{aC}), we get the corresponding cubic terms
\bea
\la{V}
S_{ssv}= \frac{4N^2}{(2\pi )^5}V_{I_1I_2I_3}\int \sqrt{-g_{a}}~
s^{I_1}\n^a s^{I_2}V_a^{I_3},
\eea
where
\bea
\la{phi13}
&&V_{I_1I_2I_3}(A)=
\frac{2(k_3+1)(\a_3-1/2)(\S-1)(\S+1)(\S+3)}
{(k_1+1)(k_2+1)(k_3+2)}t_{123}\\
\la{phi14}
&&V_{I_1I_2I_3}(C)=
\frac{8(k_3+3)(\a_3-1/2)(\a_3-3/2)(\a_3-5/2)(\S+3)}
{(k_1+1)(k_2+1)(k_3+2)}t_{123}
\eea
\vskip 0.4cm
{\it 6.2 {Three-point functions }}
\vskip 0.4cm
\noindent
Denote by ${\cal R}_i^{I_3}$ the operator in SYM that corresponds to 
$V_i^{I_3}$ on the gravity side. Then the three-point 
function of two scalars and a vector field is given by the integral
\bea 
\la{tpV}
\langle {\cal O}^{I_1}(\x ){\cal O}^{I_2}(\y ){\cal R}_i^{I_3}(\z )\rangle =
 \frac{8N^2}{(2\pi )^5}V_{I_1I_2I_3}\int \frac{d^5\omega}{\omega_0^5}
\omega_0^2 {\cal K}_{\D_1} (\o,\x)\p_{b}{\cal K}_{\D_2}(\o,\y)
G_{bi}^{I_3}(\omega,\z).
\eea
Here  ${\cal K}_{\D} (\o,\x)$ with $\D =k$ is a bulk-to-boundary propagator 
(\ref{skprop}) for $s^{I}$
and $G_{a i}(\o , \x)$ is a bulk-to-boundary propagator 
for a massive vector field $V_a^{I_3}$ with a mass $m(V)$:
\bea
\nonumber
G_{a i}(\o , \x)=\frac{\D_v}{\D_v-1}\o_0^{-1}
{\cal K}_{\D_v}(\o,\x )J_{a i}(\o-\x ),
\eea
where $J_{ab}(x)=\d_{ab}-2\frac{x_a x_b}{x^2}$. 

\noindent In the last formula 
$\D_v=2+\sqrt{1+m^2(V)}$ and, thus $\D_v=k+2$ for the field $A_a^I$
and $\D_v=k+6$ for $C_a^I$. Note that $G_{a i}$ obeys
the transversality condition $\n^a G_{a i}=0$.

The condition of the conformal covariance defines the correlator
(\ref{tpV}) uniquely up to the coefficient $\lambda_{123}$: 
\bea 
\la{tpV1}
&&\langle {\cal O}^{I_1}(\x ){\cal O}^{I_2}(\y ){\cal R}_i^{I_3}(\z )\rangle 
=\\
&&\frac{\lambda_{123}}
{|\x-\y|^{\D_1+\D_2-\D_v}|\x-\z|^{\D_1+\D_v-\D_2} |\y-\z|^{\D_2+\D_v-\D_1}}
\l \frac{|\x-\z||\y-\z|}{|\x-\y|} Z_i \r,
\nonumber
\eea
with
$$
Z_i=\frac{(\x-\z)_i}{(\x-\z)^2}-\frac{(\y-\z)_i}{(\y-\z)^2}.
$$
Applying the inversion method of  \cite{FMMR} to
integrate (\ref{tpV}) one finds for $\lambda_{123}$ the following answer
 \bea
\nonumber
\lambda_{123}&=&\frac{8N^2}{(2\pi )^5}\frac{1}{\pi^4}V_{I_1I_2I_3}
\frac{(\D_v-2)\G\l \frac{1}{2} (\D_1+\D_2+\D_v-3) \r }{\G(\D_v) } \\
\nonumber
& \times &
\frac{\G\l\frac{1}{2} (\D_1+\D_v-\D_2+1) \r
\G\l\frac{1}{2} (\D_2+\D_v-\D_1+1) \r
\G\l\frac{1}{2} (\D_2+\D_2-\D_v+1) \r  }
{ \G(\D_1-2)\G(\D_2-2) }
\eea
For the field $A^I$ the last formula reads as
\bea
\nonumber
\lambda_{123}(A)&=&\frac{4N^2}{(2\pi )^5}\frac{2^5}{\pi^4}
\frac{\G\l \frac{1}{2}\S+\frac{5}{2}\r }{(k_1+1)(k_2+1)(k_3+2)}
\frac{\G\l\a_1+3/2 \r
\G\l \a_2+3/2\r
\G\l \a_3+1/2\r  }
{ \G(k_1-2)\G(k_2-2)\G(k_3) } t_{123}
\eea
while for $C^I$:
\bea
\nonumber
\lambda_{123}(C)&=&\frac{4N^2}{(2\pi )^5}\frac{2^5}{\pi^4}
\frac{\G\l \frac{1}{2}\S+\frac{5}{2}\r (k_3+3)(k_3+4)}{(k_1+1)(k_2+1)(k_3+2)}
\frac{\G\l\a_1+7/2 \r
\G\l \a_2+7/2\r
\G\l \a_3+1/2\r  }
{ \G(k_1-2)\G(k_2-2)\G(k_3+6) } t_{123}
\eea
The two-point correlator corresponding to a massive vector field
on the AdS space was found in \cite{MV}: 
\bea
\langle {\cal R}_{i}^I(\x), {\cal R}_j^J(\y)\rangle 
=\frac{2}{\pi^2}\theta
\D_v (\D_v-1)^2 \frac{\d^{IJ}}{|\x-\y|^{2\D_v}}J_{ij}(\x-\y),
\eea
where the constant $\theta$ accounts
our normalization of the quadratic action for the 
vector fields and is equal to
$\theta=\frac{4N^2}{(2\pi )^5}\frac{k+1}{2(k+2)}$ for the field $A$ 
and to 
$\theta=\frac{4N^2}{(2\pi )^5}\frac{k+3}{2(k+2)}$ for $C$ respectively.
We introduce a normalized operator $R_i^I$ with the two-point
correlation function
$$
\langle R_{i}^I(\x), R_j^J(\y)\rangle 
=\frac{\d^{IJ}}{|\x-\y|^{2\D_v}}J_{ij}(\x-\y).
$$
Explicitly  $R_i^I$ is given by 
$$
R_i^I=\frac{(2\pi )^{5/2}}{2N}\frac{\pi}{(k+1)^{3/2}}{\cal R}_{i}^I
$$
for the YM operator corresponding to $A_a^I$ and 
$$
R_i^I=\frac{(2\pi )^{5/2}}{2N}\frac{\pi}{(k+5)}\l 
\frac{k+2}{(k+3)(k+6)} \r^{1/2}{\cal R}_{i}^I
$$
for $C_a^I$.
By using these formulae, the definition (\ref{nCPO}) of the normalized 
CPO, and the expression for $t_{123}$ from the Appendix, one gets
the correlation functions of normalized operators
\bea
\nonumber
\lambda_{123}^{norm}(A)&=&\frac{(2\pi )^{5/2}}{N}
\frac{1}{4\pi^{5/2}}
\l \frac{k_1k_2}{k_3+2}\r^{1/2}
\frac{k_3(\a_1 +1/2)(\a_2 +1/2)}{(k_3+1)^2}T_{123}
\eea
\bea
\nonumber
\lambda^{norm}_{123}(C)&=&\frac{(2\pi )^{5/2}}{N}
\frac{1}{4\pi^{5/2}}
\l \frac{k_1k_2(k_3+3)}{(k_3+1)(k_3+6)}\r^{1/2}
\frac{k_3+4}{k_3+5}\\
\nonumber
&\times&\frac{k_3!\G (\a_1+7/2)\G (\a_2+7/2)}
{\G (k_3+6)(\a_1-1/2)!(\a_2-1/2)!}T_{123}
\eea
\section{Conclusion}
In this paper we obtained the cubic couplings in type IIB 
supergravity on $AdS_5\times S^5$ involving two scalar fields
$s^I$ and the corresponding three-point functions by using
the covariant equations of motion and the quadratic action.
Since all the fields we considered correspond to operators
which are descendants of CPOs, one may, in principle, derive
the same results directly from the superconformal invariance.
This would require a detailed study of superconformal Ward identities
in SYM$_4$ which, to our knowledge, has not been carried out yet.

In most cases to find a cubic coupling of two scalars $s^I$
with a field $F$ we used a corrected equation of motion for 
the field $F$. We saw that to get rid of higher-derivative
terms we had to make a field redefinition of the form 
(\ref{red}). By this reason, a field $F$ corresponds not 
to a descendant of a CPO, but to a properly extended
operator which includes products of CPOs.

In fact one can obtain the cubic couplings
by using corrected equations of motion for scalars $s^I$ 
as it was done in the case of the graviton couplings. 
We have done that to derive the cubic couplings
of two scalars $s^I$ with the scalars $\phi^I$, 
and with the vector fields, and we have certainly obtained the 
same results (\ref{ssphi}) and (\ref{V}). The fact that we derived
the same vertices from different equations also confirms the correct
normalization of the quadratic action for type IIB supergravity \cite{AF3}.
It is worth noting that
contrary to the graviton case considered in section 5, in these
cases to remove higher-derivative terms from the corrected
equations of motion for $s^I$ we had to make the following 
redefinitions of the scalars $s^I$ 
$$
s_1\to s_1 +J_{123}s_2\phi_3 +L_{123}\n_a s_2\n^a\phi_3$$
and
$$s_1\to s_1 +J_{123}\n_a s_2V^a_3 +
L_{123}\n_a\n_b s_2\n^b V^a_3.$$
This implies that the extended CPOs corresponding to the scalars $s^I$
that were discussed in section 2 have to depend on products of CPOs
and  their descendants. 
Unfortunately, the knowledge of the three-point functions obtained 
in the paper does not allow one to fix the explicit form of
the extended CPOs uniquely. 

It is worth noting that the cubic couplings of three scalars $s$ vanish 
when any of the $\a 's$ vanish \cite{LMRS}. 
The cubic couplings studied in this 
paper vanish if
$\a_3$ takes special values, and in most of the cases there are 
several such values of $\a_3$. Since $\a_1$ and $\a_2$ have to 
be non-negative the cubic couplings have no zeroes at $\a_1$ and $\a_2$. 
However, in all of the three-point functions considered  
zeroes of the cubic couplings are cancelled by poles in 
the general expressions for 
the three-point functions, just as in the case of the three-point
functions of extended CPOs. This gives us a reason to believe that
for generic values of conformal dimensions the three-point functions
obtained coincide with the three-point functions of CPOs and
their descendants.

The next natural step is to find quartic couplings of scalars $s^I$,
and to compute four-point functions of extended CPOs. We expect
that the quartic couplings vanish if, say, $k_4=k_1+k_2+k_3$,
because in this case there is no exchange diagram, and all 
contributions to the four-point functions may be given only by 
the quartic couplings. However, the four-point functions in this case
are nonsingular at $\vec{x}_1=\vec{x_2}$, and it seems to be 
impossible to reproduce such a coordinate dependence via
supergravity with a nonzero on-shell quartic coupling.

Finally, it would be interesting to find the supergravity fields
that correspond to CPOs, and to compute their cubic couplings. 
A similar problem exists in the case of AdS compactifications
of 11-dimensional supergravity, where analogous cubic couplings 
\cite{CFMc,BZ} also have zeroes. In the 11-dimensional case the problem
seems to be simpler because the covariant action is known.  

 \vskip 1cm
{\bf ACKNOWLEDGMENT} We would like to thank Prof. S.Theisen and
Prof. J.Wess for kind hospitality at the University 
of Munich. We are grateful to Prof. S.Theisen and to S.Kuzenko
for valuable discussions and to F.Bastianelli and R.Zucchini
for pointing out the factor 2 in the Green functions
missed in the first version of the paper.
The work of G.A. was
supported by the Alexander von Humboldt Foundation and in part by the
RFBI grant N96-01-00608, and the work of S.F. was supported by
the U.S. Department of Energy under grant No. DE-FG02-96ER40967
and in part by the Alexander von Humboldt Foundation.

\section{Appendix}
We follow \cite{LMRS} describing spherical harmonics on $S^5$. 
The scalar spherical harmonics $Y^I$ are defined by
\bea
\la{ssh}
Y^I =z(k)^{-1/2}C^{I}_{i_1...i_k}x^{i_1}\cdots x^{i_k}
\eea
where $C^I_{i_1\cdots i_k}$ are totally symmetric traceless rank $k$ 
orthonormal tensors of $SO(6)$: 
$\langle C^IC^J\rangle =C^I_{i_1\cdots i_k}C^J_{i_1\cdots i_k}=\delta^{IJ}$,
$x^i$ are the Cartesian coordinates of the ${\bf R}^6$ in which $S^5$ 
is embedded, and 
$$z(k)=\frac{\pi^3}{2^{k-1}(k+1)(k+2)}$$
The scalar spherical harmonics are orthonormal and 
satisfy the relation
\bea
\la{ssh2}
&&\int~ Y^{I_1}Y^{I_2}Y^{I_3}=a_{123}
\\
\nonumber
&&a_{123}=(z(k_1)z(k_2)z(k_3))^{-1/2}
\frac{\pi^3}{(\frac12\S +2)!2^{\frac12 (\S -2)}}
\frac{k_1!k_2!k_3!}{\a_1!\a_2!\a_3!}\langle C^{I_1}C^{I_2}C^{I_3}\rangle ,
\eea
where $\a_i =\frac 12 (k_j+k_l-k_i)$, $j\neq l\neq i$, and 
$\langle C^{I_1}C^{I_2}C^{I_3}\rangle $ is the unique $SO(6)$ invariant obtained 
by contracting $\a_1$ indices between $C^{I_2}$ and  $C^{I_3}$, 
$\a_2$ indices between $C^{I_3}$ and  $C^{I_1}$, and
$\a_3$ indices between $C^{I_2}$ and  $C^{I_1}$.

A vector spherical harmonic is defined as a tangent component of the 
following vector
\bea
Y_m^I=z(k)^{-1/2}C^{I}_{m;i_1...i_k}x^{i_1}\cdots x^{i_k}
\la{defvh}
\eea
where the tensor $C^{I}_{m;i_1...i_k}$ is symmetric and traceless 
with respect to $i_1,...,i_k$, and its symmetric part vanishes. The 
tensors are orthonormal
$$
C^{I}_{m;i_1...i_k}C^{J}_{n;i_1...i_k}=\d^{IJ}\d_{mn}
$$
The vector spherical harmonics are orthonormal
and satisfy the relation
\bea
&&\int~ \n^\a Y^{I_1}Y^{I_2}Y_\a^{I_3}=t_{123}\nonumber\\
&&t_{123}=
\frac{\pi^3}{k_3+1}\frac{(z(k_1)z(k_2)z(k_3))^{-1/2}}
{(\frac12 (\Sigma +3))!
2^{\frac12 (\Sigma -3)}}\frac{k_1!k_2!k_3!}{(\a_1-\frac12)!
(\a_2-\frac12)!(\a_3-\frac12)!}T_{123}
\la{t123}
\eea
where
\bea
T_{123}=&&C^{I_1}_{mi_1...i_{p_2}j_1...j_{p_3}}
C^{I_2}_{j_1...j_{p_3}l_1...l_{p_1}}
C^{I_3}_{m;l_1...l_{p_1}i_1...i_{p_2}}-\nonumber\\
&&C^{I_1}_{i_1...i_{p_2+1}j_1...j_{p_3}}
C^{I_2}_{j_1...j_{p_3}l_1...l_{p_1-1}m}
C^{I_3}_{m;l_1...l_{p_1-1}i_1...i_{p_2+1}}
\la{v123}
\eea
and $p_1=\a_1 +\frac12$, $p_2=\a_2-\frac12$, $p_3=\a_3-\frac12$.

A tensor spherical harmonic is defined as a projection of the 
following six-dimensional tensor onto the sphere.
\bea
Y_{mn}=z(k)^{-1/2}C^{I}_{mn;i_1...i_k}x^{i_1}\cdots x^{i_k},
\la{deftenh}
\eea
where the tensor $C^{I}_{mn;i_1...i_k}$ is symmetric and traceless 
with respect to $i_1,...,i_k$, and $m,n$, and its symmetric part 
vanishes, i.e.
$$
C^{I}_{mn;i_1...i_k}+C^{I}_{mi_1;n...i_k}+\cdots +
C^{I}_{mi_k;i_1...i_{k-1}n}=0
$$
The tensors are orthonormal
$$
C^{I}_{m_1n_1;i_1...i_k}C^{J}_{m_2n_2;i_1...i_k}=\d^{IJ}\d_{m_1n_1;m_2n_2}
$$
Then, we get that the tensor spherical harmonics are orthonormal
and satisfy the relation
\bea
&&\int~ \n^\a Y^{I_1}\n^\b Y^{I_2}Y_{(\a\b )}^{I_3}=p_{123}\nonumber\\
&&p_{123}=(z(k_1)z(k_2)z(k_3))^{-1/2}
\frac{\pi^3}{(\frac12\Sigma +1)!
2^{\frac12\Sigma }}\cdot\frac{k_1!k_2!k_3!}{\a_1 !
\a_2 !(\a_3-1)!}P_{123},
\la{s123}
\eea
where
\bea
P_{123}=C^{I_1}_{mi_1...i_{p_2}j_1...j_{p_3}}
C^{I_2}_{nj_1...j_{p_3}l_1...l_{p_1}}
C^{I_3}_{mn;l_1...l_{p_1}i_1...i_{p_2}}
\la{S123}
\eea
and $p_1=\a_1$, $p_2=\a_2$, $p_3=\a_3-1$.

In deriving the equations of motions for scalar fields $t_k$
and for tensor $\phi_{(ab)}^k$ one comes across a number 
of integrals of scalar spherical harmonics, all of them can be reduced 
to $a_{123}$. Introducing the concise notation $f(k)=k(k+4)$
we present below the corresponding formulae:
\bea
\nonumber
\int \n^{\a}Y^{I_1}Y^{I_2}\n_{\a}Y^{I_3}
=\frac{1}{2}(f(k_1)+f(k_3)-f(k_2))a_{123}, 
\eea
\bea
\nonumber
\int \n^{(\a}\n^{\b)}Y^{I_1}\n_{\a}Y^{I_2}\n_{\b}Y^{I_3}&=&
\l\frac{1}{10}f(k_1)f(k_2)+\frac{1}{10}f(k_1)f(k_3)+\frac{1}{2}f(k_2)f(k_3)
\right. 
\\
\nonumber
&-&\left.\frac{1}{4}f(k_2)^2-\frac{1}{4}f(k_3)^2+\frac{3}{20}f(k_1)^2 \r
a_{123},
\eea
\bea
\nonumber
\int \n^{(\a}\n^{\b)}Y^{I_1}Y^{I_2}\n_{\a}\n_{\b}Y^{I_3}&=& 
\frac{1}{2}\l -f(k_1)f(k_2)-f(k_2)f(k_3)+\frac{3}{5}f(k_1)f(k_3)
+\frac{1}{2}f(k_1)^2 \right. \\
\nonumber
&+&\frac{1}{2}\left. f(k_2)^2+\frac{1}{2}f(k_3)^2 
-4(f(k_1)+f(k_3)-f(k_2))
\r a_{123}.
\eea

Analogously, when computing the interaction vertex $S_{ssv}$ 
from equations of motion for scalars $s_k$ one finds two 
integrals involving the vector harmonics $Y_{\a}^I$. Both of them are 
expressed via $t_{123}$: 
\bea
\nonumber
\int \n^{(\a}\n^{\b)}Y^{I_1}Y^{I_2}\n_{\a}Y_{\b}^{I_3}&=&
\frac{1}{2}\l (k_3+1)(k_3+3)-8+f(k_1)-f(k_2) \r t_{123} \\
\nonumber
\int \n^{(\a}\n^{\b)}Y^{I_1}\n_{\a}Y^{I_2}Y_{\b}^{I_3}
&=&\frac{1}{2}\l f(k_2)+\frac{3}{5}f(k_1) - (k_3+1)(k_3+3)  \r t_{123}.
\eea

Finally the derivation of the $S_{ss\phi}$-vertex from the equations 
of motion for scalars $s_k$ requires the knowledge of the following 
integrals:
\bea
\nonumber
\int \n^{(\a}\n^{\g)}Y^{I_1}\n^{\b}\n_{\g} Y^{I_2}Y_{(\a\b)}^{I_3}&=&
\frac{1}{10}(3f(k_1)+5f(k_2)-5k_3^2-20k_3-30)p_{123} \\
\nonumber
\int \n^{(\a}\n^{\b)}Y^{I_1}\n_{\g} Y^{I_2}\n^{\g}Y_{(\a\b)}^{I_3}&=&
\frac{1}{2}(f(k_1)-f(k_2)-k_3^2-4k_3-8) p_{123}.
\eea
\newpage
 
\end{document}